\documentclass[12pt,preprint]{aastex}

\newcommand{\w}{\,$\lambda$\,}
\newcommand{\ww}{\,$\lambda\lambda$\,}

\newcommand{\I}{\,{\sc i}}
\newcommand{\II}{\,{\sc ii}}
\newcommand{\III}{\,{\sc iii}}

\shorttitle{A late WN star in NGC 300}
\shortauthors{Bresolin et al.}
\slugcomment{}

\begin{document}

\title{Discovery and quantitative spectral analysis of an Ofpe/WN9
(WN11) star in the Sculptor spiral galaxy
NGC~300\footnotemark}\footnotetext[1]{Based on observations obtained
at the ESO Very Large Telescope [65.N-0389(B) and 67.D-0006(A)].}

\author{Fabio Bresolin}
\affil{Institute for Astronomy, 2680 Woodlawn Drive, Honolulu HI
96822}
\email{bresolin@ifa.hawaii.edu}

\author{Rolf-Peter Kudritzki}
\affil{Institute for Astronomy, 2680 Woodlawn Drive, Honolulu HI
96822} \email{kud@ifa.hawaii.edu}

\author{Francisco Najarro}
\affil{Instituto de Estructura de la Materia, CSIC, Serrano 121, 28006
Madrid, Spain} \email{najarro@isis.iem.csic.es}

\author{Wolfgang Gieren}
\affil{Universidad de Concepci{\'o}n, Departamento de Fisica, Casilla 160-C,
Concepci{\'o}n, Chile} \email{wgieren@coma.cfm.udec.cl}

\and

\author{Grzegorz Pietrzy{\'n}ski}
\affil{Universidad de Concepci{\'o}n, Departamento de Fisica, Casilla 160-C,
Concepci{\'o}n, Chile and\\ Warsaw University Observatory, Al.
Ujazdowskie 4, 00-478, Warsaw, Poland}
 \email{pietrzyn@hubble.cfm.udec.cl}

\begin{abstract}
We have discovered an Ofpe/WN9 (WN11 following Smith et al.) star in
the Sculptor spiral galaxy NGC~300, the first object of this class
found outside the Local Group, during a recent spectroscopic survey of
blue supergiant stars obtained at the ESO VLT. The light curve over a
five-month period in late 1999 displays a variability at the 0.1 mag
level. The intermediate resolution spectra (3800-7200~\AA) show a very
close resemblance to the Galactic LBV AG Car during minimum.  We have
performed a detailed non-LTE analysis of the stellar spectrum, and
have derived a chemical abundance pattern which includes H, He, C, N,
O, Al, Si and Fe, in addition to the stellar and wind parameters. The
derived stellar properties and the He and N surface enrichments are
consistent with those of other Local Group WN11 stars in the
literature, suggesting a similar quiescent or post-LBV evolutionary
status.

\end{abstract}


\section{Introduction}

With the new telescopes of the 8-10 meter class stellar astronomy is
branching out beyond the Local Group. Ideal targets for our
understanding of young stellar populations in distant galaxies are hot
massive stars. These objects have strong stellar winds producing broad
and easily detectable spectral features distributed over the whole
wavelength range from the UV to the IR, and providing unique
information on chemical composition, galactic evolution and
extragalactic distances. With these perspectives in mind we have
recently begun a systematic spectroscopic study of luminous blue stars
in galaxies beyond the Local Group, and presented spectral
classification and first quantitative results for A supergiants in
NGC~3621 (6.7~Mpc, \citealt{bresolin01}) and NGC~300 (2.0~Mpc,
\citealt{bresolin02}). Here we report on the discovery and detailed
quantitative analysis of the first example of an Ofpe/WN9 star outside
of the Local Group. We will present a detailed chemical abundance
pattern -- the first in a galaxy beyond the Local Group -- together
with stellar parameters and a determination of the stellar wind
properties.

The Ofpe/WN9 class was introduced to include objects which show in
their spectra high excitation emission lines from He\II\/ and N\III,
typical of Of stars, together with low excitation lines from He\I\/
and N\II, seen in late WN stars (\citealt{walborn82},
\citealt{bohannan89}). Objects of this class have so far been
identified in the Galaxy (possibly several stars in the Galactic
center: \citealt{allen90}, \citealt{najarro97b},
\citealt{figer99}), the LMC (ten stars: \citealt{bohannan89}), M33 (seven
stars: \citealt{massey96}, \citealt{crowther97b}) and M31 (one star:
\citealt{massey98}).  The importance of Ofpe/WN9 stars as objects in a
transitional stage of evolution between O and W-R stars has been
recognized in the last decade, and a connection to the LBV class has
been suggested, Ofpe/WN9 stars being observed during a quiescent or
post-LBV phase (\citealt{crowther97}, \citealt{pasquali97}). Indeed,
in at least a couple of instances Ofpe/WN9 stars have been observed to
turn into LBVs (R127: \citealt{stahl83}; HDE 269582:
\citealt{bohannan89b}).  The LBV AG Car is also known to show an
Ofpe/WN9-like spectrum during its hot phase at visual minimum
(\citealt{stahl86}).  The discovery of ejected circumstellar nebulae,
in some cases measured in a state of expansion, associated with some
of the LMC Ofpe/WN9 stars by
\citet{nota96} and \citet{pasquali99} brings forward strong evidence
for the occurence of violent episodes of mass loss in these stars,
similar to the shell-producing, eruptive outbursts of LBVs.

\citet{smith94} revised and extended the classification of late WN 
stars to include lower excitation objects.  In their scheme, stars
showing spectra like AG Car at minimum, where no N\III\/ is detected,
are reclassified as WN11. Given the spectral resemblance of the NGC
300 star we analyze here to AG Car at visual minimum (see \S~4) we
will adopt the WN11 classification for the remainder of this Letter.
We describe the observational data in
\S~2, and the photometry in \S~3. In \S~4 we present our VLT spectra,
together with the stellar and wind properties derived from a
quantitative spectral analysis.

\section{Observations}
The data presented here are part of a spectroscopic survey of
photometrically selected blue supergiants in the Sculptor spiral
galaxy NGC~300 obtained at the VLT with the FORS multiobject
spectrograph, and described in detail by \citet{bresolin02}. This
latter paper presents spectra obtained in September 2000 in the blue
spectral region ($\sim$\,4000-5000~\AA) and used for the spectral
classification of about 70 supergiants. The emission-line star
analyzed here corresponds to star B-16 of the spectral catalog (see
Table~2 and finding charts in the aforementioned paper), a rather
isolated bright star apparently not associated with any prominent OB
association, star cluster or nebulosity.  The nearest OB association
(from the catalog of \citealt{pietrzynski01}) lies at a deprojected
distance of $\sim$450 pc. Given the estimated age of 3.8 Myr (\S
4.1) a speed on the order of 100 km/s would be required for the star
to reach its current position from this association. This is much
larger than the typical stellar random velocities, and comparable to that
of Galactic runaway stars.  To the best of our knowledge the star is
not included in published listings of emission-line or W-R stars in
NGC~300 (\citealt{schild92}, \citealt{breysacher97}, and references
therein), and we will refer to it as star B-16.

In a more recent observing run (September 2001), in order to measure
the mass-loss rates of blue supergiants in NGC~300 from the H$\alpha$
line profile, we secured MOS spectra in the red with a 600 lines\,mm$^{-1}$
grating and 1\arcsec\/ slitlets, which, together with the older
spectra, provides us with complete coverage at $R\simeq1,000$
resolution of the 3800-7200~\AA\/ wavelength range, except for a
50~\AA-wide gap centered at 5025~\AA.  The total exposure time was
13,500\,s for both the blue and the red spectra, while better seeing
conditions favored the blue observations. The mass-loss rates we
derive for the whole blue supergiant sample will be discussed in a
forthcoming paper, and we concentrate here on the single B-16 star.

\section{Photometry}
\begin{figure}
\plotone{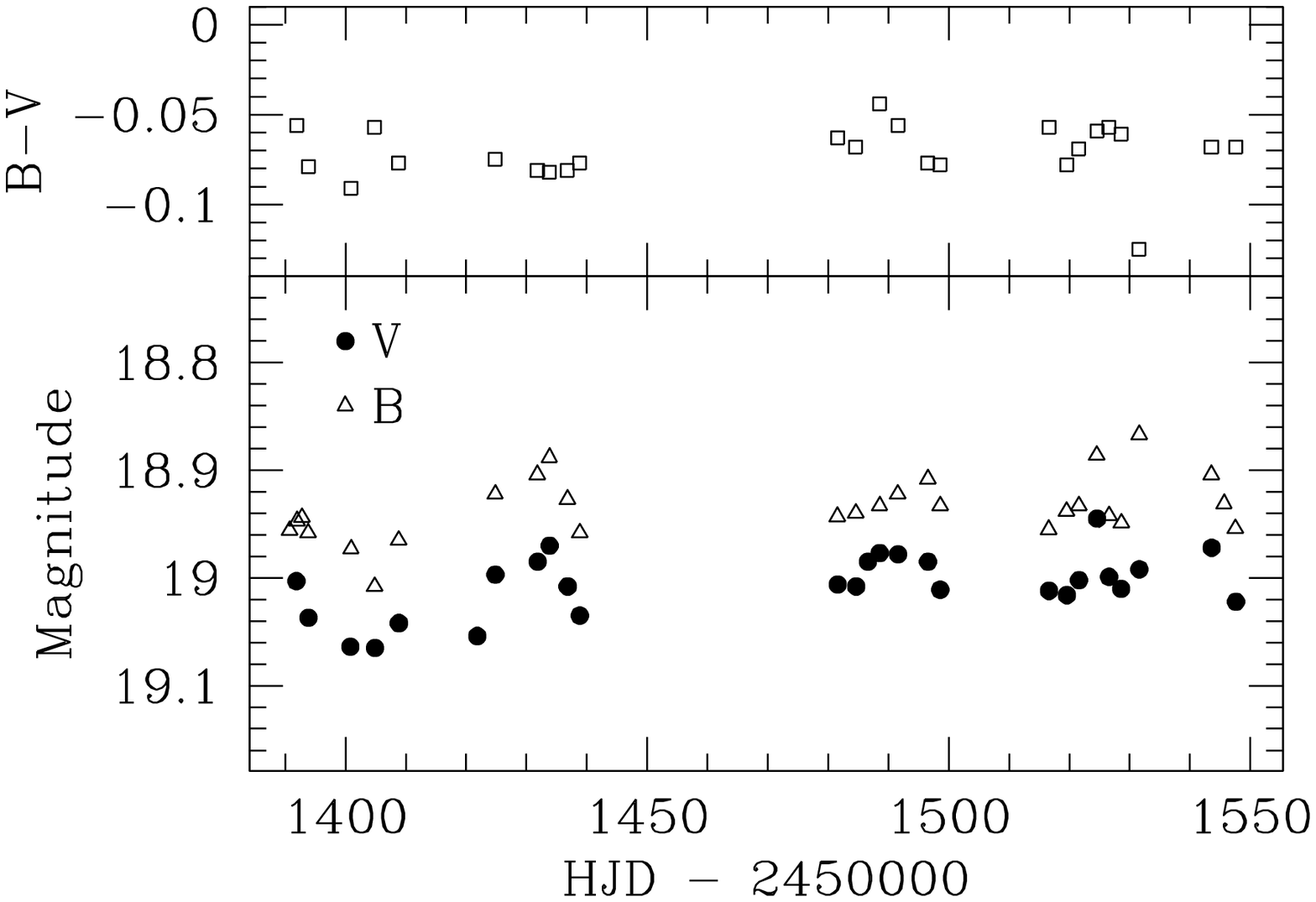}
\caption{$B$ and $V$ light curves for NGC~300 star B-16, spanning an
approximately five month period between September 1999 and January
2000. The $B-V$ color index is shown in the top panel, the $V$ (full
dots) and $B$ (open triangles) magnitudes in the bottom panel.
Photometric errors ($\sim$0.015 mag) are comparable to the size of the
dots. \label{curve}}
\end{figure}

As part of a multi-epoch photometric study of NGC~300 carried out for
the discovery and monitoring of Cepheid variables
(\citealt{pietrzynski01}, see also \citealt{pietrzynski02}), $BV$
photometry from ESO/MPI 2.2m telescope WFI images is available for
B-16 covering nearly a one-half year period. The resulting $B$ and $V$
light curves are shown in Fig.~\ref{curve}. An apparently irregular
variability, on the order of 0.1 mag (peak-to-peak), is detected in
both bands, at a significant level above the typical photometric
uncertainty of 0.015 mag. The color index remains approximately
constant around $B-V=-0.07$.  While {\em bona fide} LBV variables show
large amplitude variability on timescales of years
(\citealt{stahl01}), partly characterizing the outburst phenomena in
these objects, smaller-amplitude variability is observed in Ofpe/WN9
stars or LBVs during quiescence (e.g., \citealt{stahl84},
\citealt{sterken97}).  Assuming the intrinsic color index from the
stellar atmosphere models ($E_{B-V}=0.06$), $R=A_V/E_{B-V}=3.1$ and an
average magnitude $V=19.00$, we obtain from the adopted distance
modulus for NGC~300, $m-M=26.53$ (\citealt{freedman01}), an absolute
magnitude $M_V\simeq-7.72\pm0.13$.
 
\section{Spectral analysis}
\begin{figure}
\plotone{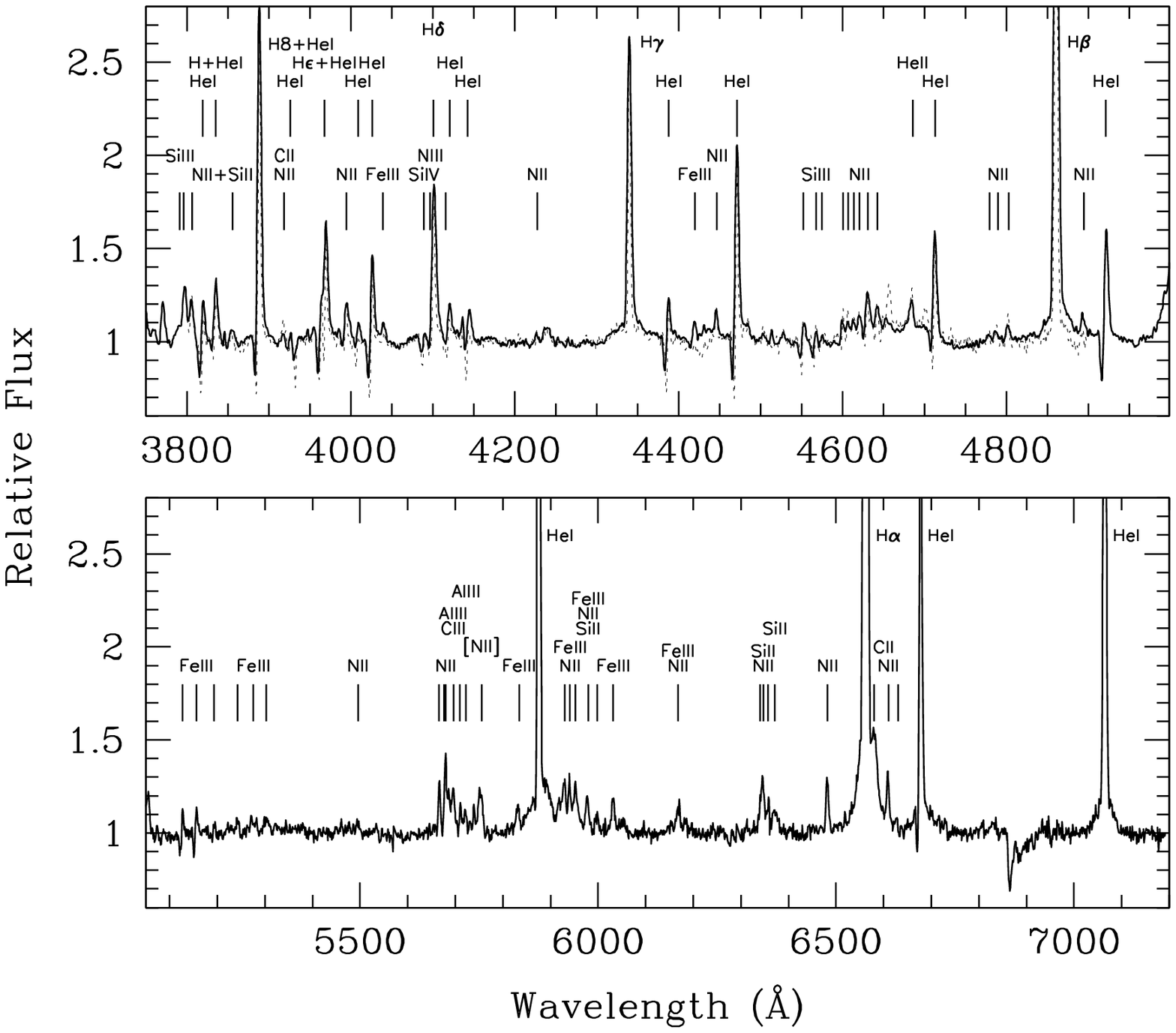}
\caption{Rectified VLT/FORS optical spectra obtained at blue (top) and red
(bottom) wavelengths. Line identification is provided in the top
section of each panel. The comparison spectrum of AG Car at visual
minimum (courtesy N. Walborn and E. Fitzpatrick) is shown as a dashed
line superimposed on top of the blue spectrum of star B-16.
\label{spectrum}}
\end{figure}

The blue and red portions of the rectified VLT optical spectrum are
shown in Fig.~\ref{spectrum}. In the top panel we have also
superimposed the spectrum of AG Car during minimum, taken from the
\citet{walborn00} atlas, in order to illustrate the remarkable
similarity between the two spectra. Line identification is provided
for most of the recognizable features, which characterize B-16 as a
WN11 star.  These include the hydrogen Balmer series lines (pure
emission), He\I\/ lines (together with He\II\w4686) and mostly
low-excitation metal lines such as N\II, Fe\III\/ and Si\II. P-Cygni
profiles are seen in most of the He\I\/ lines, as well as in
Si\III\ww4552-4575, N\II\ww4601-4643 and Fe\III\ww5127,5156. The
absence of N\III\/ is a discriminant factor against an earlier
(WN9-10) classification. The WN11 classification is further confirmed
by the observed equivalent widths of He\I\w5876 (32\,\AA) and
He\II\w4686 (1\,\AA), compared to those given by \citet{crowther97}.

\subsection{Atmosphere models}

For the quantitative analysis of the stellar spectrum we have used the
iterative, non-LTE line blanketing method presented by
\citet{hillier98}. The code solves the radiative transfer equation in
the co-moving frame for the expanding atmospheres of early-type stars
in spherical geometry, subject to the constraints of statistical and
radiative equilibrium.  Steady state is assumed, and the density
structure is set by the mass-loss rate and the velocity field via the
equation of continuity.  The velocity field (\citealt{hillier89}) is
characterized by an isothermal effective scale height in the inner
atmosphere, and becomes a $\beta$ law in the wind (e.g.,
\citealt{lamers96}).  Better fits are obtained if instead of the
standard $\beta$ law a combined 2-$\beta$ law is used, where each
value of $\beta$ (given in Table~\ref{observations}) characterizes the
shape of the velocity field in the inner and outer parts of the wind,
respectively.  We allow for the presence of clumping via a clumping
law characterized by a volume filling factor $f(r)$, so that the
`smooth' mass-loss rate, $\dot{M}_S$, is related to the `clumped'
mass-loss rate, $\dot{M}_C$, through $\dot{M}_S =
\dot{M}_C/f^{1/2}$.  The model is then prescribed by the stellar
radius $R_*$, the stellar luminosity $L_*$, the mass-loss rate
$\dot{M}$, the velocity field $v(r)$, the volume filling factor $f$, and
the abundances of the elements considered.  The reader is referred to
\citet{hillier98,hillier99} for a detailed discussion of the code.

\begin{figure}
\plotone{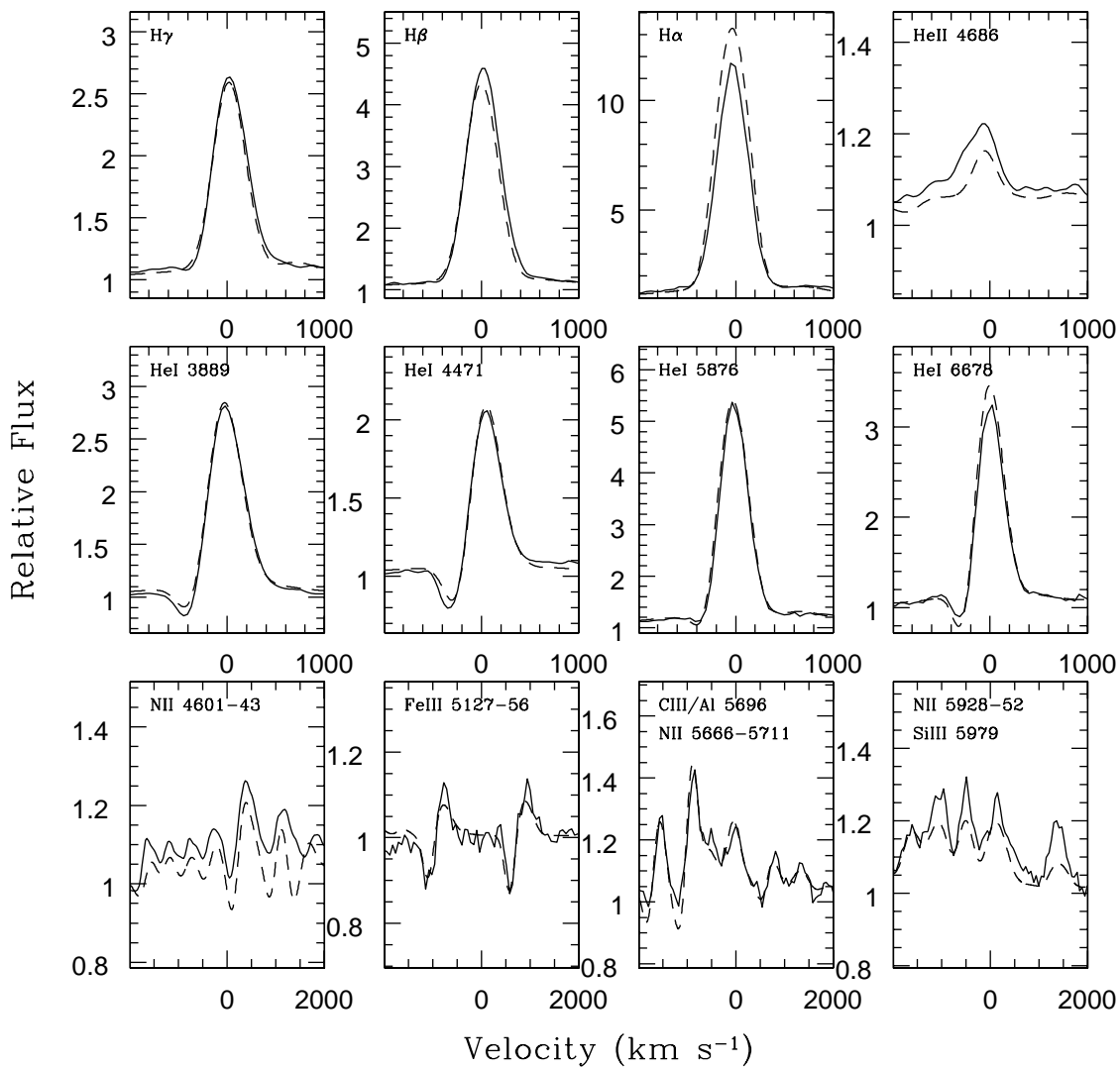}
\caption{Different spectral diagnostics in the observed spectrum of
B-16 compared to our best fit model (dashed line), convolved with a
Gaussian profile to match the spectral resolution of the data.
\label{model}}
\end{figure}

\begin{figure}
\plotone{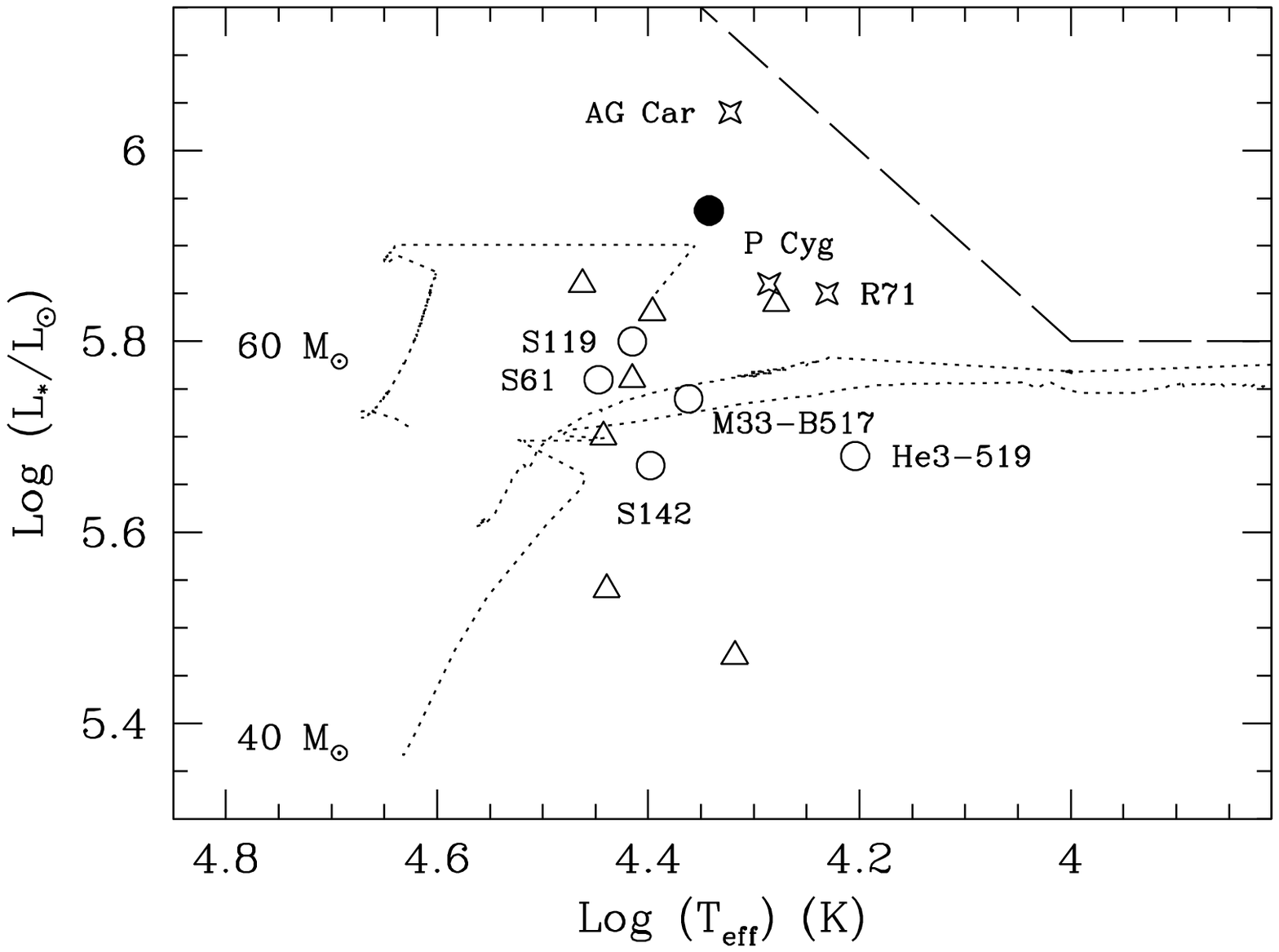}
\caption{Location of B-16 in the H-R diagram (full circle), plotted
together with additional WN11 stars (open circles, labeled) and WN9-10
stars (open triangles; from Crowther et al.~1995; Smith et al.~1995;
\citealt{crowther97b}), as well as LBVs at minimum (open stars,
labeled) in the Milky Way (P Cyg: \citealt{pauldrach90}; AG Car:
\citealt{smith94}) and in the LMC (R71: \citealt{lennon94}). We
include the Humphreys-Davidson limit (dashed line), and the
theoretical Geneva stellar tracks including rotation ($v_{ini}=300$
km\,s$^{-1}$) at solar metallicity, for $M=40~M_\odot$ and
$60~M_\odot$ (dotted lines). These tracks are plotted from the
zero-age main sequence to the point where the fractional mass of
hydrogen $X\simeq0.20$.
\label{hr}}
\end{figure}

The main stellar parameters of our best-fitting model are summarized
in Table~\ref{observations}, whereas fits to some of the most
important line diagnostics are illustrated in Fig.~\ref{model}. As can
be seen, the majority of the spectral features are well reproduced by
the adopted model. The Fe\III\ww5127,5156 lines, together with
He\II\w4686, provide excellent constraints for the effective
temperature. Using the strength of the electron scattering wings of
the Balmer lines we derive a clumping factor $f=0.25$. Thus the
"smooth" mass loss rate would be twice the value displayed in
Table~\ref{observations}. The parameters describing the stellar wind,
i.e., the terminal velocity $v_\infty=325$~km\,s$^{-1}$, the mass-loss
rate $\dot{M_C}=4.6\times10^{-5}$~$M_\odot$\,yr$^{-1}$, and the wind
performance number $\eta=\dot{M}v_\infty/(L_*/c)=0.84$, lie in the
range found for Local Group WN11 stars (\citealt{crowther97b}).  We
estimate the uncertainty in $L$ and $\dot{M_C}/f^{1/2}$ to be
approximately 10\% and 30\%, respectively.  As a further comparison,
Fig.~\ref{hr} shows the location of star B-16 in the H-R diagram,
together with additional WN9-11 stars in the Local Group
(\citealt{crowther95}; \citealt{smith95}; \citealt{crowther97b}), as
well as LBVs at minimum in the Galaxy (P Cyg: \citealt{pauldrach90};
AG Car: \citealt{smith94}) and in the LMC (R71: \citealt{lennon94}).

Table~\ref{abundances} summarizes the model fractional mass
abundances. Uncertainties in the latter range between 0.2 dex (He, N,
Si and Fe) and 0.3 dex (C, O and Al).  The chemistry of B-16 resembles
that of other known Local Group WN9-11 stars.  The severe depletion of
hydrogen inferred from our analysis (mass fraction $X=0.27$), and the
correspondingly large helium surface abundance (H/He~=~1.5 by number),
are fairly typical for late WN stars (H/He~=~0.8-3 by number), and are
indicative of heavy mass-loss stripping of the stellar outer layers
during the post-main sequence evolution of massive stars. We find a
total CNO abundance close to that of Galactic main sequence B stars
(\citealt{gies92}, \citealt{kilian92}), consistent with a half solar
chemical composition in these elements. This agrees with the empirical
O abundance derived for an H~{\sc ii} region located 30\arcsec\/ away
(region 5 observed by \citealt{pagel79}). On the other hand, Fe has a
roughly solar abundance, suggesting an $\alpha$/Fe ratio different
than in the Galaxy. In comparison with the Galactic B star abundance
pattern, N is overabundant (by mass) by a factor of 10, while C and O
are reduced by factors of $\sim$2 and 6, respectively, although the
latter abundances are highly uncertain.

The current evolutionary status of B-16 can be estimated by means of
the recent Geneva stellar tracks including rotation
(\citealt{meynet00}) with an initial rotational velocity
$v_{ini}=300$~km\,s$^{-1}$ and Z~=~0.02 (Fig.~\ref{hr}).  The star's
position in the H-R diagram approaches the 60~$M_\odot$ track, and an
initial mass of $\sim$55~$M_\odot$ can also be derived from the
relation between stellar luminosity and initial mass for WNL stars
given by \citet{schaerer92}. From the 60~$M_\odot$ stellar model more
closely matching the observed H and He surface abundances we infer an
age of 3.8 Myr and a present mass of $\sim$36~$M_\odot$.  At this
stage, the predicted N overabundance with respect to the ZAMS
abundance is 12 (mostly occuring during the MS phase, as a consequence
of rotational mixing), in good agreement with our finding.

According to its He- and N-enriched surface chemistry, B-16 might be
in a dormant or post-LBV phase of evolution. This is also indicated by
the position in the H-R diagram, well above the Humphreys-Davidson
limit (\citealt{humphreys79}), and intermediate between the LBVs
AG~Car and P Cygni, which have comparable mass-loss rates to B-16, as
well as similar wind performance numbers.  The NGC~300 star shows
higher He enrichment at the surface (n$_{He}$/n$_H\simeq0.7$, versus
0.4-0.5 for the LBVs), and a somewhat faster wind (325 versus
185-250~km\,s$^{-1}$; \citealt{najarro97}, \citealt{langer94}), so
that a more advanced (post-LBV) state seems more likely.  This picture
is in agreement with the conclusions reached by \citet{crowther95} and
\citet{crowther97b} regarding the close connection between WN9-11
stars and LBVs.

An unexpected reward out of our spectroscopic survey of blue
supergiants in NGC~300, the WN11 star B-16 studied here is the first
star in this galaxy (and beyond the Local Group) for which we have
attempted a detailed quantitative analysis. In the near future the
systematic study of the massive stellar population in NGC~300 and
similar galaxies within a few Mpc, well within current observational
capabilities, will certainly provide new insights into massive stellar
evolution and stellar abundances.

\acknowledgments We thank J. Hillier for providing his code, and
G. Meynet for the Geneva stellar evolutionary tracks including
rotation.  WG gratefully acknowledges support for this research from
the Chilean Center for Astrophysics FONDAP No. 15010003. FN
acknowlegdes Spanish MCYT PANAYA2000-1784 and Ramon y Cajal grants.


\epsscale{1.0}

\epsscale{1.0}

\epsscale{0.9}

\epsscale{1.0}

\begin{deluxetable}{lr}
\tabletypesize{\scriptsize}
\tablecolumns{2}
\tablewidth{0pt}
\tablenum{1}
\tablecaption{Basic properties and model parameters\label{observations}}
\tablehead{}
\startdata 
R.A.\phantom{.} (J2000)	&	\phantom{$-$}00 54 44.64 	\\
Decl. (J2000)	&	$-$37 42 39.13 	\\
$V$ (mean)		&	19.00		\\
$B-V$ (mean)		&	$-0.07$		\\
$M_V$		&	$-7.72$		\\

\sidehead{\em Model fit parameters}
$E_{B-V}$	&	0.06	\\
$\log(L_*/L_\odot)$\tablenotemark{a}	&	5.94	\\	
$T_*$ (K)\tablenotemark{a}	&	26000	\\
$T_{\rm eff}$ (K)\tablenotemark{b} &	22000	\\
$R_*/R_{\odot}$\tablenotemark{a}	  &	46	\\
$R_{2/3}/R_{\odot}$\tablenotemark{b}	  &	64	\\
$\log(\dot{M_C})$ ($M_\odot\,yr^{-1}$)  &	$-4.34$ \\
$v_\infty$ ($km\,s^{-1}$)	&	325 \\
$\beta_1$			&	6.0 \\
$\beta_2$			&	0.9 \\
$f$				&	0.25 \\
$\dot{M_C}v_\infty/(L_*/c)$	&	0.84
\enddata
\tablenotetext{a}{$\tau\simeq20$}
\tablenotetext{b}{$\tau\simeq2/3$}
\end{deluxetable}

\begin{deluxetable}{cccc}
\tabletypesize{\scriptsize}
\tablecolumns{4}
\tablewidth{0pt}
\tablenum{2}
\tablecaption{Model abundances\label{abundances}}
\tablehead{
\colhead{Species}	& 
\colhead{Relative number}	&
\colhead{Mass}		&
\colhead{$X/X_\odot$}\\
\colhead{}		&
\colhead{fraction}	&
\colhead{fraction}		&
\colhead{}}
\startdata 
H	&	1.5\,E$+$00 & 2.7\,E$-$01	&	0.4 \\
He	&	1.0\,E$+$00 & 7.2\,E$-$01	&	2.7 \\
C	&	3.6\,E$-$04 & 7.7\,E$-$04	&	0.3 \\
N	&	2.5\,E$-$03 & 6.3\,E$-$03	&	7.7 \\
O	&	5.5\,E$-$04 & 1.6\,E$-$03	&	0.2 \\
Al	&	5.0\,E$-$06 & 2.4\,E$-$05	&	0.4 \\
Si	&	8.5\,E$-$05 & 1.7\,E$-$04	&	0.2 \\
Fe	&	1.5\,E$-$04 & 1.5\,E$-$03	&	1.3 \\
\enddata
\end{deluxetable}


\end{document}